\documentclass[nohyper]{JHEP3}

\usepackage{amsmath,amssymb}
\usepackage[dvips]{graphicx}
\usepackage{afterpage}
\usepackage{pifont}
\usepackage{array}

\newcommand{\deltacp}{\ensuremath{\delta_\text{CP}}}
\newcommand{\stheta}{\ensuremath{\sin^2 2\theta_{13}}}
\newcommand{\phidet}{\phi^\text{det}}
\newcommand{\hphidet}{{\hat\phi}^\text{det}}
\newcommand{\Br}{\text{Br}}

\newcommand{\ie}{{\it i.e.}}
\newcommand{\eg}{{\it e.g.}}
\newcommand{\cf}{{\it cf.}}
\newcommand{\etc}{{\it etc.}}
\newcommand{\Eq}{Eq.}

\newcommand{\Fig}{Fig.}
\newcommand{\Figs}{Figs.}
\newcommand{\Ref}{Ref.}
\newcommand{\Refs}{Refs.}
\newcommand{\Sec}{Sec.}

\newcommand{\App}{Appendix}
\newcommand{\Tab}{Table}

\newcommand{\equ}[1]{\Eq~\eqref{equ:#1}}
\newcommand{\figu}[1]{\Fig~\ref{fig:#1}}

\newcommand{\PAGEFIGURE}[1]{\FIGURE[!p]{#1}\afterpage\clearpage}

\newcommand{\auxNH}[3]{%
  \raisebox{-3.7mm}[0.0mm][5.0mm]{\makebox[0pt][l]{\ding{#1}}}%
  \raisebox{-0.8mm}[0.0mm][0.0mm]{\makebox[0pt][l]{\ding{#2}}}%
  \raisebox{+3.7mm}[7.5mm][0mm]{\makebox{\ding{#3}}}}

\newcommand{\auxIH}[3]{%
  \raisebox{-3.7mm}[0.0mm][5.0mm]{\makebox[0pt][l]{\ding{#1}}}%
  \raisebox{+0.8mm}[0.0mm][0.0mm]{\makebox[0pt][l]{\ding{#2}}}%
  \raisebox{+3.7mm}[7.5mm][0.0mm]{\makebox{\ding{#3}}}}

\newcommand{\auxSL}[1]{\makebox[0pt][l]{\hspace{#1}\large/}}

\newcommand{\dgN}[1]{%
  \ifnum#1=1\auxNH{182}{183}{184}\else%
  \ifnum#1=2\auxNH{182}{183}{174}\else%
  \ifnum#1=3\auxNH{182}{173}{184}\else%
  \ifnum#1=4\auxNH{172}{183}{184}\else%
  \ifnum#1=5\auxNH{172}{173}{184}\else%
  \ifnum#1=6\auxNH{172}{183}{174}\else%
  \ifnum#1=7\auxNH{182}{173}{174}\else%
  \ifnum#1=8\auxNH{172}{173}{174}\fi\fi\fi\fi\fi\fi\fi\fi}

\newcommand{\dgI}[1]{%
  \ifnum#1=1\auxIH{184}{182}{183}\else%
  \ifnum#1=2\auxIH{184}{182}{173}\else%
  \ifnum#1=3\auxIH{184}{172}{183}\else%
  \ifnum#1=4\auxIH{174}{182}{183}\else%
  \ifnum#1=5\auxIH{174}{172}{183}\else%
  \ifnum#1=6\auxIH{174}{182}{173}\else%
  \ifnum#1=7\auxIH{184}{172}{173}\else%
  \ifnum#1=8\auxIH{174}{172}{173}\fi\fi\fi\fi\fi\fi\fi\fi}

\newcommand{\mdN}[1]{%
  \ifnum#1=1{LMH}\else%
  \ifnum#1=2{LM\auxSL{1pt}H}\else%
  \ifnum#1=3{L\auxSL{1pt}MH}\else%
  \ifnum#1=4{\auxSL{0pt}LMH}\else%
  \ifnum#1=5{\auxSL{0pt}L\auxSL{1pt}MH}\else%
  \ifnum#1=6{\auxSL{0pt}LM\auxSL{1pt}H}\else%
  \ifnum#1=7{L\auxSL{1pt}M\auxSL{1pt}H}\else%
  \ifnum#1=8{\auxSL{0pt}L\auxSL{1pt}M\auxSL{1pt}H}\fi\fi\fi\fi\fi\fi\fi\fi}

\newcommand{\mdS}[1]{%
  \ifnum#1=1{123}\else%
  \ifnum#1=2{12\auxSL{-1pt}3}\else%
  \ifnum#1=3{1\auxSL{-0.5pt}23}\else%
  \ifnum#1=4{\auxSL{-0.7pt}123}\else%
  \ifnum#1=5{\auxSL{-0.7pt}1\auxSL{-0.5pt}23}\else%
  \ifnum#1=6{\auxSL{-0.7pt}12\auxSL{-1pt}3}\else%
  \ifnum#1=7{1\auxSL{-0.5pt}2\auxSL{-1pt}3}\else%
  \ifnum#1=8{\auxSL{-0.7pt}1\auxSL{-0.5pt}2\auxSL{-1pt}3}\fi\fi\fi\fi\fi\fi\fi\fi}

\preprint{IFT-UAM/CSIC-08-16}

\title{Testing neutrino flavor mixing plus decay with neutrino
  telescopes}

\author{Michele Maltoni\\
  Departamento de F\'isica Te\'orica \& Instituto de F\'isica
  Te\'orica UAM/CSIC, Facultad de Ciencias C-XI, Universidad
  Aut\'onoma de Madrid, Cantoblanco, E-28049 Madrid, Spain\\
  E-mail: \email{michele.maltoni@uam.es}}

\author{Walter Winter\\
  Institut f\"ur theoretische Physik und Astrophysik,
  Universit\"at W\"urzburg, \\ D-97074 W\"urzburg, Germany\\
  E-mail: \email{winter@physik.uni-wuerzburg.de}}

\abstract{%
  We discuss the interplay of neutrino oscillation and decay
  properties at neutrino telescopes. Motivated by recent unparticle
  scenarios, which open the possibility of new neutrino decay modes
  over astrophysical distances, we perform a complete classification
  of possible decay schemes, and we illustrate how different scenarios
  can be identified. Moreover, we show that the sensitivity of
  neutrino telescopes to standard neutrino properties, such as the
  mass hierarchy or $\deltacp$, is greatly enhanced in specific decay
  scenarios. In particular we discuss the impact of an astrophysical
  neutrino detection on terrestrial experiments, such as on the mass
  hierarchy measurement at NO$\nu$A. For example, we find that the
  scenario where only $\nu_1$ is stable can be uniquely identified
  against all the other decay schemes, and that in this case CP
  violation can be established (for large $\theta_{13}$) by the
  combination of Double Chooz with the track-to-shower ratio at a
  neutrino telescope, even if the flavor composition at the source is
  unknown. Our statements are based on a complete analysis of all the
  present solar, atmospheric, reactor and accelerator neutrino data,
  as well as on realistic simulation of future terrestrial neutrino
  oscillation experiments.}

\keywords{Neutrino flavor mixing, neutrino decay, neutrino telescopes}

\begin{document}

\section{Introduction}

Neutrino telescopes~\cite{Aslanides:1999vq, Ahrens:2002dv,
Tzamarias:2003wd, Piattelli:2005hz} are sensitive to neutrinos with an
average energy and traveled distance many orders of magnitude larger
than present neutrino experiments, and provide therefore a completely
new window on both standard and non-standard neutrino properties.
Apart from ``conventional'' parameters, such as neutrino masses and
mixing angles, a very prominent example of such properties is the
neutrino lifetime. Phenomenologically, from the observation of
neutrinos from supernova 1987A, we know that at least one neutrino
mass eigenstate must be stable over galactic distances.
More stringent and explicit bounds can be derived from different
observations when specific decay models are assumed (see, \eg,
\Refs~\cite{Pakvasa:2003db, Yao:2006px, Pakvasa:2008nx} for an
overview). For example, solar neutrinos strongly limit the possibility
of radiative decays~\cite{Raffelt:1985rj}, while for Majoron
decays~\cite{Gelmini:1980re, Chikashige:1980qk} explicit bounds can be
obtained from neutrinoless double-beta decay and
supernovae~\cite{Tomas:2001dh}. Purely phenomenological (\ie,
model-independent) bounds are, however, much weaker, leaving enough
parameter space for the decay of any mass eigenstate over
extragalactic distances~\cite{Joshipura:2002fb, Bandyopadhyay:2002qg,
GonzalezGarcia:2008ru}. 
Neutrino decays are usually described by a factor $\exp[-t /
(\tau\gamma)] = \exp[-(L m_i) / (E \tau_i)]$, where $\tau_i$ is the
rest frame lifetime of the $\nu_i$ mass eigenstate boosted by $\gamma
= E/m_i$ in the laboratory frame. Since $L$ and $E$ are given by the
experiment and $m_i$ is unknown, one typically quotes $\tau_i/m_i$ as
the neutrino lifetime. The best direct limit is obtained for $\nu_1$
from SN1987A to be about $\tau_1/m_1 \gtrsim
10^5~\text{s/eV}$~\cite{Hirata:1987hu}. Bounds on $\nu_2$ lifetime are
imposed by solar neutrino data, yielding $\tau_2/m_2 \gtrsim
10^{-4}~\text{s/eV}$ for invisible decays~\cite{Joshipura:2002fb,
Bandyopadhyay:2002qg} and $\tau_2/m_2 \gtrsim 10^{-3}~\text{s/eV}$ for
decay modes with secondary $\bar{\nu}_e$
appearance~\cite{Eguchi:2003gg, Aharmim:2004uf}. Finally, limits on
$\nu_3$ follow from the analysis of atmospheric and long-baseline
neutrino data, $\tau_3/m_3 \gtrsim
10^{-10}~\text{s/eV}$~\cite{GonzalezGarcia:2008ru}. Neutrino
telescopes can probe lifetimes as long as $\sim 10^7~\text{s/eV} \cdot
L \, \text{[kpc]} \,/\, E \, \text{[TeV]}$, which are about a factor
of $10^5$ longer than the current best limit (for $E \sim
1~\text{TeV}$). 
In view of the rather weak direct neutrino lifetime limits
and the recently proposed unparticle
models, which may lead to new mechanisms of neutrino
decay~\cite{Chen:2007zy, Zhou:2007zq, Li:2007kj, Majumdar:2007mp}, we
do not assume any specific decay model, but study the most general
case. Note that neutrino telescopes may also probe different kinds of
new physics (see, \eg, \Refs~\cite{Pakvasa:2008nx, Awasthi:2007az,
Quigg:2008ab}), which we do not discuss in this work.

In addition to decay properties, the propagation from the neutrino
source to the detector is described by the neutrino mixing parameters
through averaged neutrino oscillations, \ie, by neutrino flavor mixing. 
If the neutrino telescope has
some flavor identification capability, this dependence can be used to
extract information on the decay~\cite{Beacom:2002vi, Lipari:2007su, Majumdar:2007mp}
and oscillation~\cite{Farzan:2002ct, Serpico:2005sz, Serpico:2005bs,
Bhattacharjee:2005nh, Winter:2006ce, Majumdar:2006px, Blum:2007ie,
Rodejohann:2006qq, Xing:2006xd, Hwang:2007na, Choubey:2008di}
parameters, in a way which might be synergistic to terrestrial
measurements. For example, reactor neutrino experiments in combination
with astrophysical observations might provide hints on $\deltacp$ well
before superbeams, by measuring the CP-even part of the oscillation
probabilities~\cite{Winter:2006ce, Blum:2007ie}. In addition, the
sensitivity to the oscillation parameters can be enhanced in some
decay scenarios~\cite{Beacom:2003zg}. In this study, we discuss the
identification of the various neutrino properties in scenarios with
both neutrino flavor mixing and decay.  Since uncertainties in the
mixing parameters~\cite{Meloni:2006gv} and flavor composition at
the source~\cite{Pakvasa:2007dc} may limit such measurements, we
carefully include these aspects in our study.

\textbf{Astrophysical sources and flavor composition.}
The existence of astrophysical neutrinos is not yet proven, but the
detection of very high energy cosmic rays points towards cosmic
accelerators which are expected to produce in addition high energy
neutrinos. There are many potential candidates for neutrino sources,
such as gamma ray bursts, active galactic nuclei or starburst
galaxies, the latter being unaffected by the Waxmann-Bahcall
bound~\cite{Loeb:2006tw}. Astrophysical neutrinos are normally assumed
to originate from pion decays, with a flavor ratio at the source of
$(f_e,\, f_\mu,\, f_\tau) \simeq (1/3,\, 2/3,\, 0)$ arising from the
decays of both primary pions and secondary muons (``pion beam
source''); here $f_\alpha$ is the fraction of flavor $\nu_\alpha$
(neutrinos and antineutrinos combined), so that $f_e + f_\mu + f_\tau
= 1$. However, it was pointed out in \Ref~\cite{Kashti:2005qa} that
such sources may become opaque to muons at higher energies, in which
case the flavor ratio at the source changes to $(f_e,\, f_\mu,\,
f_\tau) \simeq (0,\, 1,\, 0)$ (``muon damped source''). Therefore, one
can expect a smooth transition from one type of source to the other as
a function of the neutrino energy~\cite{Kachelriess:2006fi,
Kachelriess:2007tr}. Once a specific neutrino source is found and
identified, for example from its energy spectrum or using information
from its optical counterpart, it might be possible to select a
specific flavor ratio at the source by applying suitable energy cuts
to the data. However, note that, since the neutrino flux drops as the
energy increases, we can expect less events from muon damped sources
than from pion beam sources.
In this work we will mainly focus on pion beam and on muon damped
sources, as well as on neutrinos produced by photo-dissociation of
heavy nuclei with a flavor composition $(f_e,\, f_\mu,\, f_\tau)
\simeq (1,\, 0,\, 0)$ at the source~\cite{Anchordoqui:2003vc,
Hooper:2004xr} (``neutron beam source''). In some cases we will
consider arbitrary flavor compositions at the source without
significant tau neutrino production, as is expected from a diffuse
flux coming from the superposition of pion beam and muon damped
sources with different energy dependencies. For the identification of
even more generalized sources including the possibility of tau
neutrinos, see \eg\ \Ref~\cite{Xing:2006uk}. 

\textbf{Detector and observables.}
On the detector side, flavor identification is the prerequisite to
learn about neutrino properties. In a neutrino telescope such as
IceCube, muon tracks can be most easily seen for $E \gtrsim 100 \,
\text{GeV}$. Electron and tau neutrinos will produce showers with a
somewhat higher energy threshold, $E \gtrsim 1 \, \text{TeV}$. In
general, it is not possible to distinguish between electron and tau
events close to the threshold, whereas at much higher energies one may
be able to identify these flavors as well~\cite{Beacom:2003nh}. In
particular, one may use the ``double-bang'' signature of $\nu_\tau$ in
a window $5 \cdot 10^{14} \, \text{eV} \lesssim E \lesssim 2 \cdot
10^{16} \, \text{eV}$ to distinguish all
flavors~\cite{Learned:1994wg}. It is therefore plausible to assume
that the ratios $R \equiv \phidet_\mu / (\phidet_e + \phidet_\tau)$
(tracks/showers) and $S \equiv \phidet_e / \phidet_\tau$
(electromagnetic/hadronic showers) can be used as observables, where
$\phidet_\alpha = \hphidet_{\alpha^+} + \hphidet_{\alpha^-}$ is the
neutrino (+) plus antineutrino (-) flux of flavor $\nu_\alpha$ at the
detector~\cite{Winter:2006ce, Blum:2007ie}. As an additional
observable, one may use $T \equiv \hphidet_{e^-} / \phidet_\mu$ for
the Glashow resonance process $\bar{\nu}_e + e^- \to W^- \to
\text{anything}$ at around $6.3 \, \text{PeV}$~\cite{Learned:1994wg,
Anchordoqui:2004eb, Bhattacharjee:2005nh} to distinguish
between neutrinos and antineutrinos. Note that $T$ is detectable only
in a very narrow energy range, and that due to the difficulties in the
identification of the double-bang the precision on $S$ will be lower
than the one on $R$. Moreover, for muon damped sources $S$ will only
be measurable in rare cases, because the typical energy window of this
kind of source coincides only occasionally with the double-bang
window.

This work is organized as follows. In \Sec~\ref{sec:physics} we
describe the considered physics scenarios and we present a general
classification of all possible decay schemes. In
\Sec~\ref{sec:decayid} we focus on normal mass hierarchy and discuss
the possibility of identifying the decay scenario. The impact of the
neutrino mass hierarchy and the possibility to establish it from
astrophysical pion damped sources is discussed in
\Sec~\ref{sec:hierarchy}. In \Sec~\ref{sec:gensource} we extend our
results to the case of unknown flavor compositions at the source, as
is the case for certain diffuse fluxes. In \Sec~\ref{sec:glashow} we
study the Glashow resonance process as an additional observable,
addressing the problem of the separation between neutrino and
antineutrinos. In \Sec~\ref{sec:synergies} we illustrate synergies
with terrestrial neutrino oscillation experiments.  Finally, in
\Sec~\ref{sec:summary} we summarize our results and draw conclusions.
Details of our statistics treatment can be found in
\App~\ref{app:statistics}.

\section{Considered physics scenarios}
\label{sec:physics}

\TABLE[!t]{
  \setlength{\tabcolsep}{2pt}
  \begin{tabular}{llm{48mm}m{23mm}|c|c|c|c|c|c|c|c|}
      &&&& \$1 & \$2 & \$3 & \$4 & \$5 & \$6 & \$7 & \$8
      \\
      && Branchings ratios && \mdS1 & \mdS2 & \mdS3 & \mdS4 &
      \mdS5 & \mdS6 & \mdS7 & \mdS8
      \\
      \hline
      \#1 & \mdN1 & -- & --
      & \dgN1\,\dgI1 & -- & -- & -- & -- & -- & -- & --
      \\
      \hline
      \#2 & \mdN2
      & $\Br_\mathrm{H \to M} = a$, $\Br_\mathrm{H \to L} = b$
      \par $\Br_\mathrm{H \to I} = 1-a-b$
      & $0 \le a \le 1$\par $0 \le b \le 1-a$
      & -- & \dgN2 & \dgI2 & -- & -- & -- & -- & --
      \\
      \hline
      \#3 & \mdN3
      & $\Br_\mathrm{M \to L} = a$, $\Br_\mathrm{M \to I} = 1-a$
      & $0 \le a \le 1$
      & -- & -- & \dgN3 & \dgI3 & -- & -- & -- & --
      \\
      \hline
      \#4 & \mdN4
      & $\Br_\mathrm{L \to I} = 1$ & --
      & -- & \dgI4 & -- & \dgN4 & -- & -- & -- & --
      \\
      \hline
      \#5 & \mdN5
      & $\Br_\mathrm{M \to I} = 1$\par $\Br_\mathrm{L \to I} = 1$
      & --
      & -- & -- & -- & -- & \dgN5 & \dgI5 & -- & --
      \\
      \hline
      \#6 & \mdN6
      & $\Br_\mathrm{H \to M} = a$, $\Br_\mathrm{H \to I} = 1-a$
      \par $\Br_\mathrm{L \to I} = 1$
      & $0 \le a \le 1$
      & -- & -- & -- & -- & -- & \dgN6 & \dgI6 & --
      \\
      \hline
      \#7 & \mdN7
      & $\Br_\mathrm{H \to L} = a$, $\Br_\mathrm{H \to I} = 1-a$
      \par $\Br_\mathrm{M \to L} = b$, $\Br_\mathrm{M \to I} = 1-b$
      & $0 \le a \le 1$\par $0 \le b \le 1$
      & -- & -- & -- & -- & \dgI7 & -- & \dgN7 & --
      \\
      \hline
      \#8 & \mdN8
      & \multicolumn{2}{l|}{Not relevant, since no neutrinos observed}
      & -- & -- & -- & -- & -- & -- & -- & \dgN8\,\dgI8
      \\
      \hline
  \end{tabular}
  \caption{\label{tab:phys}%
    Classification of all the possible decay scenarios for complete
    decays, according to both LMH (rows) and 123 (columns) naming
    conventions. The tags ``L'', ``M'', ``H'' refer to the lightest,
    middle, heaviest active mass eigenstate, respectively, whereas
    ``I'' refers to an invisible state. The tags ``1'', ``2'', ``3''
    refer to the $\nu_1$, $\nu_2$, $\nu_3$ active mass eigenstates. A
    slash through the tag means that the corresponding state is
    unstable. The icons illustrate the correspondence between the two
    naming conventions according to the given mass hierarchy. The
    black and white disks correspond to stable and unstable mass
    eigenstates, respectively.}}

In this work we consider the most general combination between neutrino
flavor mixing and arbitrary neutrino decay scenarios. This includes the
conventional picture of flavor mixing among stable states as a limiting
case. Concerning the oscillation part, we assume that flavor mixing
take place only among the three known neutrino flavors, which means
that sterile neutrino states --~if they exist~-- do not mix with the
active ones. As for the decay part, following the approach of
\Ref~\cite{Beacom:2002vi} we assume that all unstable mass eigenstates
have decayed between the source and the detector, \ie, the decays are
\emph{complete}. Moreover, we neglect possible differences between
neutrino and antineutrino decay rates, and we assume that if one
polarity has completely decayed, the other one has as well. The decays
products may be \emph{visible} to the neutrino detector, \ie,
different active states, or \emph{invisible} for the detector, such as
sterile neutrinos, unparticle states, Majorons \etc. Since the decay
is assumed to be complete and neutrino oscillations are completely
averaged over astrophysical distances, the transition probabilities
are independent of the neutrino energy. Therefore, in this study we do
not take into account possible information on the energy spectrum, and
only focus on total rates. In particular, we assume that the daughter
neutrinos, if active states, fully contribute to the observed signal
regardless of whether they are degraded in energy. Note that in some
scenarios interference effects between oscillations and decay may
occur if the source is coherent and the neutrinos decay while they are
still oscillating~\cite{Lindner:2001fx, Lindner:2001th}; however, for
the sake of simplicity we do not consider such cases or the
corresponding corrections. 

Let us now consider all possible decay scenarios in a systematic way.
First of all, note that for kinematical reasons any mass eigenstate
can only decay into lighter ones, which in turn may be stable or
unstable. However, the assumption of \emph{complete} decay allows to
eliminate intermediate unstable states from every decay chain. For
example, if the heaviest eigenstate decays into both the middle and
the lightest state, and the middle state decays into the lightest
state, finally everything will end up in the lightest state. This
argumentation includes more complicated scenarios with arbitrary
branching ratios, including active states decaying into invisible
states which then decay back into active ones, as long as the initial
states are active. It means that the transition probabilities can be
written in terms of the {\em effective} branching ratios $\Br_{i \to
f}$ between the initial \emph{unstable} active states $\nu_i$ and the
final \emph{stable} active states $\nu_f$:
\begin{equation}
    \label{equ:prob}
    P_{\alpha \beta}
    = \sum_{f~\text{stable}} \hspace{-1mm} \left( |U_{\alpha f}|^2
    + \sum_{i~\text{unstable}} \hspace{-3mm} |U_{\alpha i}|^2
    \, \Br_{i \to f} \right) |U_{\beta f}|^2 \,.
\end{equation}
Note that $\sum_f \Br_{i \to f} = 1$ only if there are no invisible
final states, whereas in general $\sum_f \Br_{i \to f} \le 1$. Thus
this formula also accounts for invisible final states.

As we have seen, any decay scenario is uniquely characterized by the
stability of its active states. In general, there are $2^3=8$
possibilities, since either active state may be stable or not. We list
these possibilities in \Tab~\ref{tab:phys}, together with the relevant
parameters needed to completely describe each scenario. It is
convenient to classify the various decay scenarios according to two
different naming conventions:
\begin{description}
  \item[LMH classification.]
    This naming convention is illustrated in the \emph{rows} of
    \Tab~\ref{tab:phys}. The labels ``L'', ``M'', ``H'' refer to the
    lightest ($\nu_L$), middle ($\nu_M$), heaviest ($\nu_H$) active
    mass eigenstate, respectively; a slash through the label (\eg,
    ``\auxSL{0pt}L'', ``\auxSL{1pt}M'', ``\auxSL{1pt}H'') means that
    the corresponding state is unstable. Different scenarios are
    denoted by $\#n$, with $n=1\dots 8$.
    
  \item[123 classification.]
    This naming convention corresponds to the \emph{columns} of
    \Tab~\ref{tab:phys}. The labels ``1'', ``2'', ``3'' refer to the
    mass eigenstates $\nu_1$, $\nu_2$, $\nu_3$ relevant for neutrino
    flavor mixing, irrespective of their mass ordering; as before, a
    slash through the label (\eg, ``\auxSL{-0.7pt}1'',
    ``\auxSL{-0.5pt}2'', ``\auxSL{-1pt}3'') means that the
    corresponding state is unstable. Different scenarios are denoted
    by $\$n$, with $n=1\dots 8$.
\end{description}
The correspondence between these two classifications depends on the
neutrino mass hierarchy; it is illustrated in \Tab~\ref{tab:phys}.
Specifically, for the normal hierarchy we simply have $\#n = \$n$,
whereas for the inverted hierarchy we have $\#1 = \$1$, $\#2 = \$3$,
$\#3 = \$4$, $\#4 = \$2$, $\#5 = \$6$, $\#6 = \$7$, $\#7 = \$5$ and
$\#8 = \$8$. Note that the branching ratios $\Br_{I \to F}$ are
completely independent from the neutrino mass hierarchy when $I$ and
$F$ are written in the LMH notation. Similarly, the matrix elements
$|U_{\alpha i}|^2$ and $|U_{\alpha f}|^2$, which appear in \equ{prob},
are insensitive to the mass hierarchy when $i$ and $j$ are 123 tags.
On the other hand, combinations of \emph{both} mixing angles and
branching ratios, such as the transition probabilities
$P_{\alpha\beta}$ in \equ{prob}, do depend on the mass hierarchy. We
will discuss the impact of the neutrino mass hierarchy in greater
detail in \Sec~\ref{sec:hierarchy}.

The observables $R$ and $S$ defined in the previous section can be
computed as
\begin{equation}
    \label{equ:flr}
    R = \frac{\sum_\alpha f_\alpha \, P_{\alpha \mu}}
    {\sum_\alpha f_\alpha \, (P_{\alpha e} + P_{\alpha \tau})} \,,
    \qquad
    S = \frac{\sum_\alpha f_\alpha \, P_{\alpha e}}
    {\sum_\alpha f_\alpha \, P_{\alpha \tau}} \,.
\end{equation}
Note that the overall flux normalization, which depends on the source
luminosity, distance to the source, \etc, cancels in this definition.
In addition, although the cross sections are not very well known at
high energies, the valence quark contribution becomes negligible, and
the dependence on the flavor becomes small~\cite{Gandhi:1995tf,
Gandhi:1998ri}. Therefore, we expect that the uncorrelated cross
section error among the different flavors are small, whereas the
correlated cross section error cancels in \equ{flr}.

An interesting special case is when there is only a \emph{single}
active stable neutrino mass eigenstate $\nu_f$. Substituting
\equ{prob} in \equ{flr}, it is straightforward to see that (see also
\Ref~\cite{Pakvasa:1981ci})
\begin{equation}
    \label{equ:flrspecial}
    R = \frac{|U_{\mu f}|^2}
    {|U_{e f}|^2 + |U_{\tau f}|^2}
    = \frac{|U_{\mu f}|^2}{1 - |U_{\mu f}|^2} \,,
    \qquad
    S = \frac{|U_{e f}|^2}{|U_{\tau f}|^2}
\end{equation}
since the probabilities factorize in source-dependent and
detector-dependent parts. This means that in scenarios \#6 and \#7 the
quantities $R$ and $S$ do not depend on the parameters $a$ or $b$
listed in \Tab~\ref{tab:phys}, even if the probabilities do. In
addition, \equ{flrspecial} implies that there is no dependence on the
$f_\alpha$ characterizing the source, which means that the
uncertainties on the source flavor composition are irrelevant for
scenarios \#5, \#6 and \#7.

\FIGURE[!t]{
  \includegraphics[width=0.95\textwidth]{fig.oscparm.eps}
  \caption{\label{fig:params}%
    The observable $R$ for a pion beam source as function of
    $\deltacp$ (left) and $\sin^2 \theta_{23}$ (right) for the
    different scenarios in \Tab~\ref{tab:phys} (normal hierarchy
    assumed). We have chosen $a=b=0.5$ for scenarios \#2 and \#3.}}

To explicitly illustrate the kind of implications that the observation
of an astrophysical neutrino source could have for neutrino
phenomenology, we show in \figu{params} the dependence of $R$ on the
parameters $\deltacp$ (for large $\stheta$) and $\sin^2 \theta_{23}$
(for small $\stheta$). For definiteness we focus on the normal mass
hierarchy and a pion beam source. As soon as a particular physics
scenario is identified, a concrete measurement of $R$ may considerably
help in the determination of $\deltacp$ or $\theta_{23}$. In turn, if
$\deltacp$, $\theta_{13}$, and $\theta_{23}$ are constrained, we can
use $R$ to infer the decay scenario. For what concerns $\deltacp$,
scenarios \#1, \#2 and \#5 are obviously unfortunate, since there is
little or no dependence on this parameter. However, if nature has
implemented scenarios \#3, \#4, \#6 or \#7 astrophysical sources may
provide very important information on $\deltacp$.  As for
$\theta_{23}$, and in particular the octant determination, all
scenarios may help, but scenarios \#4, \#5, and \#6 are especially
well suited. Note that there is some parameter dependence on the
branching ratios in scenarios \#2 and \#3, while there is no
dependence on the flavor composition at the source for scenarios \#5,
\#6, and \#7 as explained above. Such synergies between astrophysical
and terrestrial experiments will be discussed in detail in the next
sections.

Sometimes it may be useful to test additional assumptions coming from
specific models. In addition to the general case, in this work we will
consider the following special cases:
\begin{description}
  \item[Special case 1:]
    the lightest mass eigenstate $\nu_L$ is stable. This constraint
    might be motivated by the observation of neutrinos from supernova
    1987A. Only scenarios \#1, \#2, \#3, and \#7 are compatible with
    this assumption.
    
  \item[Special case 2:]
    there are no invisible states. It follows that the lightest state
    must be stable, since it could only decay into a sterile state.
    Therefore, this case is a special realization of the previous one.
    In addition, the branching parameters become constrained: scenario
    \#2 has now only one parameter ($b=1-a$), and scenario \#3 has no
    parameters at all ($a=1$). 
\end{description}

\section{Physics scenario identification}
\label{sec:decayid}

\PAGEFIGURE{
  \includegraphics[width=0.95\textwidth]{fig.dblchooz.eps}
  \caption{\label{fig:flratios}%
    Allowed regions at 99\% CL in the $(R,\, S)$ plane corresponding
    to different decay scenarios, for a muon damped source (left
    panels) and a pion beam source (right panels). We assume a normal
    hierarchy. The upper panels correspond to the analysis of present
    data reported in \Ref~\cite{GonzalezGarcia:2007ib}. The other
    panels show the impact of 3 years of Double Chooz data taking (1.5
    with near detector), assuming no signal ($\stheta = 0$, middle
    panels) or a large signal ($\stheta=0.1$, lower panels). The extra
    branching ratio parameters $a$ and $b$ have been varied as well,
    where applicable.}}

Let us now focus on the normal mass hierarchy (we will discuss the
impact of the mass hierarchy in the next section) and let us assume
that we can identify the source. Furthermore, we assume that
information on both observables $R$ (muon tracks to showers) and $S$
(electromagnetic to hadronic showers) will be available. Under these
hypotheses, the measurement of an astrophysical neutrino flux
corresponds to a point in the $(R,\, S)$ plane, with certain
measurement errors.

In order to discuss to which extent one can in principle disentangle
different physics scenarios, we show in \figu{flratios} the 99\%
allowed regions corresponding to different decay scenarios.
Specifically, we project the global $\chi^2$ from present and future
terrestrial experiments onto the $(R,\, S)$ plane for each scenario, 
as the oscillation parameters and branching ratio parameters $a$ and
$b$ (where applicable) are varied. The left and right columns
correspond to muon damped and pion beam sources, respectively. In the
upper panels we show 99\% regions implied by the global analysis of
present solar, atmospheric, reactor and accelerator neutrino
data~\cite{GonzalezGarcia:2007ib}. These experiments are further
combined with the accurate measurement of $\stheta$ expected after 3
years of Double Chooz data taking (1.5 of these with near detector),
assuming that no signal ($\stheta = 0$, middle panels) or a large
signal ($\stheta=0.1$, lower panels) is observed.

Let us first of all focus on a pion beam source and current
experiments, \ie, the upper right panel in \figu{flratios}. Ignoring
the uncertainties in the astrophysical measurement, scenarios \#4, \#5
and \#7 are clearly separated from each other and from the rest.
Although there is some overlap among the other scenarios, the physics
can still be clearly identified in many case. For example, a
measurement $(R,\, S) \simeq (1,\, 1.5)$ would uniquely determine
scenario \#6. Conversely, if one measures $(R,\, S) \simeq (0.6,\, 1)$
there is some ambiguity among scenarios \#1, \#3, and \#6, but
scenarios \#2, \#4, \#5, and \#7 can still be excluded. Even if the
observable $S$ won't be available, in many cases one can clearly
identify or exclude certain scenarios using only the projection onto
the $R$-axis. For example, a measurement $R=1.5$ can only arise from
scenario \#5. The situation obviously improves if one includes in the
analysis future terrestrial experiments such as Double Chooz, since
the regions become somewhat smaller, although the qualitative picture
does not change. Note, however, that the standard oscillation scenario
\#1 can never be uniquely established from astrophysical measurements,
unless a further hypothesis on the stability of certain mass
eigenstates are assumed a priori.

If we compare the pion beam source (upper right panel) with a muon
damped source (upper left panel), we see that in general the allowed
regions are much larger for the muon damped source. This means that
the physics scenario identification becomes more difficult, but it
also implies that the dependence on the individual parameters is
stronger. Indeed it is well known that in the standard oscillation
case (\#1) the dependence of $R$ and $S$ on $\sin^2 \theta_{23}$ and
$\deltacp$ is considerably stronger for the muon damped source than
for the pion beam source. It is also clear that the information from
different sources is somewhat synergistic for what concerns the
physics scenario identification. For example, $(R,\, S) = (0.7,\,
0.5)$ single out scenario \#4 for a pion beam source, whereas the
scenario cannot be determined for a muon beam source. In turn, $(R,\,
S) = (1.2,\, 0.2)$ not only points towards scenario \#3, but also
uniquely identifies the source as muon damped, since for a pion beam
source no region is present in this point of the parameter space.
Moreover, note that scenarios \#5, \#6 and \#7, which only have one
stable active mass eigenstate, are independent of the source type, as
we have pointed out in \equ{flrspecial}.

\TABLE[!t]{
  \begin{tabular}{lrr|rrrrrrr|rr} 
      \hline
      \multicolumn{3}{l}{Simulated scenario}
      & \multicolumn{8}{c}{Fit scenario $\Delta \chi^2$} \\
      No. & $R$ & $S$ & \#1 & \#2 & \#3 & \#4 & \#5 & \#6 & \#7 & Any & $\sigma$ \\
      \hline 
      \multicolumn{10}{l}{\bf L=100, R+S measured:} \\
      \#1 & 0.49 & 1.07 &   -- &   9.0 &  0.1 &  26.4 & 220.3 &   0.6 &  55.3 &  0.1 & 0.4 \\
      \#2 & 0.38 & 2.19 & 21.0 &    -- &  0.4 &  73.1 & 432.1 &  20.0 &  20.6 &  0.4 & 0.6 \\
      \#3 & 0.38 & 1.47 &  5.1 &   0.4 &   -- &  47.4 & 274.0 &   7.7 &  32.7 &  0.4 & 0.6 \\
      \#4 & 0.69 & 0.41 & 20.7 &  53.3 & 17.9 &    -- &  43.7 &  15.4 & 115.7 & 15.4 & 3.9 \\
      \#5 & 0.83 & 0.00 & 70.3 & 100.3 & 67.0 &  29.8 &    -- &  62.9 & 153.5 & 29.8 & 5.5 \\
      \#6 & 0.59 & 1.04 &  1.1 &   8.4 &  0.9 &  12.8 & 106.2 &    -- &  48.2 &  0.9 & 1.0 \\
      \#7 & 0.21 & 4.67 & 77.6 &  11.2 & 22.3 & 138.5 & 516.2 &  54.8 &    -- & 11.2 & 3.3 \\
      \hline
      \multicolumn{10}{l}{\bf L=10, R+S measured:} \\
      \#1 & 0.49 & 1.07 &  -- &  1.2 & 0.0 &  4.8 & 36.2 &  0.1 & 13.0 & 0.0 & 0.2 \\
      \#2 & 0.38 & 2.19 & 2.6 &   -- & 0.1 & 15.0 & 59.7 &  3.8 &  4.0 & 0.1 & 0.3 \\
      \#3 & 0.38 & 1.47 & 0.6 &  0.1 &  -- &  8.2 & 41.7 &  1.2 &  6.2 & 0.1 & 0.3 \\
      \#4 & 0.69 & 0.41 & 2.8 &  6.5 & 2.9 &   -- & 10.5 &  2.1 & 20.7 & 2.1 & 1.5 \\
      \#5 & 0.83 & 0.00 & 7.7 & 10.9 & 7.8 &  3.3 &   -- &  7.0 & 20.6 & 3.3 & 1.8 \\
      \#6 & 0.59 & 1.04 & 0.1 &  1.0 & 0.2 &  2.0 & 19.8 &   -- &  7.9 & 0.1 & 0.3 \\
      \#7 & 0.21 & 4.67 & 8.7 &  1.2 & 3.2 & 23.4 & 67.7 & 10.4 &   -- & 1.2 & 1.1 \\
      \hline
      \multicolumn{10}{l}{\bf L=100, Only R measured:} \\
      \#1 & 0.49 & 1.07 &   -- & 0.3 & 0.1 &  3.8 &  4.7 & 0.1 &  7.9 & 0.1 & 0.3 \\
      \#2 & 0.38 & 2.19 &  3.7 &  -- & 0.0 & 13.6 & 11.6 & 1.4 &  2.4 & 0.0 & 0.0 \\
      \#3 & 0.38 & 1.47 &  2.8 & 0.0 &  -- & 11.2 & 10.3 & 1.2 &  2.4 & 0.0 & 0.0 \\
      \#4 & 0.69 & 0.41 &  4.8 & 4.2 & 2.9 &   -- &  0.3 & 0.3 & 18.4 & 0.3 & 0.5 \\
      \#5 & 0.83 & 0.00 &  5.7 & 5.5 & 4.3 &  0.6 &   -- & 1.0 & 16.8 & 0.6 & 0.7 \\
      \#6 & 0.59 & 1.04 &  1.1 & 1.5 & 0.9 &  0.2 &  1.3 &  -- & 10.5 & 0.2 & 0.5 \\
      \#7 & 0.21 & 4.67 & 23.3 & 1.6 & 3.6 & 37.0 & 30.0 & 9.2 &   -- & 1.6 & 1.3 \\
      \hline
  \end{tabular}
  \caption{\label{tab:ex}%
    Exclusion of different physics scenarios from present data plus
    astrophysical measurements, assuming normal mass hierarchy and a
    pion beam source. The first three columns refer to the simulated
    scenario and the simulated (benchmark) $R$ and $S$ values
    corresponding to the current best-fit values for the oscillation
    parameters and $a=b=0.5$ (where applicable). These points are
    marked in the upper right panel of \figu{flratios}. The next seven
    columns give the $\Delta \chi^2$ at which the corresponding decay
    scenario can be excluded, marginalized over all branching ratios
    and oscillation parameters. In the last two columns we marginalize
    also with respect to the fitted ``wrong'' scenarios. The different
    groups assume different statistics (muon tracks $L$) and either
    $R$ and $S$ as observables, or only $R$ (\cf,
    \App~\ref{app:statistics}). The $\chi^2$ from present solar,
    atmospheric, reactor, and accelerator data has been
    added~\cite{GonzalezGarcia:2007ib}.}}

As the next step, let us simulate a realistic astrophysical
measurement with concrete statistics. The details of our simple
simulation are given in \App~\ref{app:statistics}. Note that we
normalize the source luminosity to the number $L$ of muon tracks which
would be observed in the detector in the absence of neutrino decay
(scenario \#1) and for $\stheta=0$. This means that we are properly
using the same source luminosity for all the different scenarios.
In general, a $1\sigma$ error of order 10\% might be expected for
$\mathcal{O}(100)$ events~\cite{Winter:2006ce}. Since this error is
much smaller than the typical size of the allowed regions shown in
\figu{flratios}, we expect that our considerations hold as long as
there are enough events. This statement is quantified in
\Tab~\ref{tab:ex}, which is based on the combination of present data
with a simulated astrophysical measurement. The first three columns
refer to the simulated scenario and the simulated $R$ and $S$ values,
which are also plotted in the upper right panel of \figu{flratios}.
The next seven columns give the $\Delta \chi^2$ at which the
corresponding decay scenario can be excluded, marginalized over all
branching ratios and oscillation parameters. In the last two columns
we marginalize also with respect to the fitted scenarios, giving the
$\Delta \chi^2$ and number of sigmas at which the simulated scenario
can be established against {\em all} the others. The various blocks
refer to different assumptions about statistics (muon tracks $L$) and
observables (either $R$ alone or both $R$ and $S$ measured).
This table clearly illustrates that even for low statistics (middle
block) many scenarios can be excluded. For example, for simulated
scenario \#1 and only 10 muon tracks, scenarios \#3 and \#7 are ruled
out at more than $3\sigma$. The same conclusion holds for higher
statistics but no $S$ measurement (lower block). In the high
statistics case with both $R$ and $S$ measured (upper block) it is
possible to uniquely establish scenarios \#4, \#5, and \#7, excluding
\emph{all} the other ones.

An interesting and somewhat simpler issue is whether we can establish
the stability of the lightest neutrino, \ie, if special case~1 of
\Sec~\ref{sec:physics} is realized. This corresponds to the following
questions: if the true scenario is one of \#1, \#2, \#3, and \#7 (in
which the lightest neutrino is stable), can we rule out scenarios \#4,
\#5, and \#6? Conversely, if the real scenario is one of \#4, \#5, and
\#6 (in which the lightest neutrino is unstable), can we rule out
scenarios \#1, \#2, \#3, and \#7? Focusing again on normal hierarchy
and on a pion beam source, and assuming that both $R$ and $S$ can be
measured, we can answer these questions by looking at the upper right
panel of \figu{flratios}. Qualitatively, the perspectives to establish
the stability of the lightest neutrino state depend on the true decay
scenario, as follows:
\begin{description}
  \item[\#1:]
    never (it is contained in \#6);
    
  \item[\#2:]
    always (apart from a very small overlap with \#6), if statistics
    is good enough;
    
  \item[\#3:]
    sometimes, if no overlap with \#6 and there is enough statistics;

  \item[\#4:]
    always, if statistics is good enough;

  \item[\#5:]
    always, even with low statistics;

  \item[\#6:]
    sometimes, if no overlap with \#2 and \#3 and there is enough
    statistics;

  \item[\#7:]
    always, even with low statistics.
\end{description}
Therefore, the chances to determine whether the lightest state is
stable or not over astrophysical distances are quite high.

\section{Neutrino mass hierarchy}
\label{sec:hierarchy}

\FIGURE[!t]{
  \includegraphics[width=0.95\textwidth]{fig.hierarchy.eps}
  \caption{\label{fig:hierarchy}%
    Allowed regions at 99\% CL in the $(R,\, S)$ plane corresponding
    to different decay scenarios, for the normal hierarchy (left
    panels) and the inverted hierarchy (right panels). We assume a
    pion beam source. The upper panels show the general case where all
    the scenario are allowed, while the lower panels correspond to the
    assumption that lightest mass eigenstate is stable (Special
    Case~1, see \Sec~\ref{sec:physics}). If we further strengthen this
    restriction by imposing that there are no invisible states
    (Special Case~2), the patterned regions can be excluded as well.}}

As we have seen in \Sec~\ref{sec:physics}, the neutrino mass hierarchy
plays a crucial role in the correspondence between the two different
naming conventions introduced in \Tab~\ref{tab:phys}. Therefore, a
detailed discussion of the main features of each scheme can help to
understand the impact of the mass hierarchy on the scenario
identification discussed so far.
In the language of the LMH scheme, the branching ratios --~and
therefore the propagation from the astrophysical source to the
neutrino detector~-- are completely independent of the hierarchy. On
the other hand, the projection of the flavor states $(\nu_e,\,
\nu_\mu,\, \nu_\tau)$ onto the mass eigenstates $(\nu_L,\, \nu_M,\,
\nu_H)$ is different between normal and inverted hierarchy. This means
that the production and detection processes, when described in terms
of $(\nu_L,\, \nu_M,\, \nu_H)$, look different in the two hierarchies.
For example, for the normal hierarchy, $\nu_e$ is mostly $\nu_L$, but
for the inverted hierarchy, it is mostly $\nu_M$. Therefore, the LMH
convention provides a simpler description of the decay scenarios and
is also more motivated from a theoretical (decay model) point of view,
but it is not appropriate to describe oscillation phenomena.
On the other hand, in the language of the 123 scheme, the relation
between the flavor states $(\nu_e,\, \nu_\mu,\, \nu_\tau)$ and the
mass states $(\nu_1,\, \nu_2,\, \nu_3)$ is the same for both
hierarchies, but there is an asymmetry in the branching ratios created
by kinematics, since heavier states can only decay into lighter ones.
As an example, let us consider scenario \$2 ($\nu_1$ and $\nu_2$
stable, $\nu_3$ unstable). For the normal hierarchy, $\nu_3$ is the
heaviest state and can decay into $\nu_1$, into $\nu_2$, or into
invisible states, with a plethora of branching possibilities which are
described in terms of two parameters (\cf\ \Tab~\ref{tab:phys}).
Conversely, for the inverted hierarchy, $\nu_3$ is the lightest state
and can only decay into invisible states without any further freedom.
This implies that for scenario \$2 the branching possibilities for the
inverted hierarchy are only a subset of those for the normal
hierarchy, and hence the allowed region in the $(R,\, S)$ plane for
the inverted hierarchy is a subregion of the corresponding one for
normal hierarchy. Similar analogies can be derived for scenarios \$3
and \$4. Note that scenario \$1 does not depend on the branching
ratios and therefore there is no asymmetry between the normal and
inverted hierarchy regions. The same happens for scenarios \$5, \$6
and \$7, since in this case the expressions for $R$ and $S$ are given
by \equ{flrspecial}, which does not contain the branchings. Note,
however, that there are very small discrepancies between the two
hierarchies due to the slight asymmetry introduced by present
data~\cite{GonzalezGarcia:2007ib}.

Let us now discuss the the impact of the neutrino mass hierarchy on
the physics scenario identification. In \figu{hierarchy} we show the
allowed regions in the $(R,\, S)$ plane corresponding to different
decay scenarios, for the normal hierarchy (left panels) and the
inverted hierarchy (right panels). The colors represent the different
scenarios in the LMH classification, but each region is explicitly
labeled according to both schemes.  From this figure we observe that
the allowed domains corresponding to the same 123 scenario but to
different hierarchies are quite similar, and one of the two is always
a subregion of the other. Therefore, all the considerations presented
in \Sec~\ref{sec:decayid} about the identification of the decay
scenario in the case of normal mass hierarchy are still qualitatively
valid for the case of unknown hierarchy, provided that they are
reformulated in the language of the 123 scheme. In other words, by
measuring the $(R,\, S)$ parameters we may be able to uniquely
establish which of the $(\nu_1,\, \nu_2,\, \nu_3)$ eigenstates are
stable, but in order to convert this into a statement on the stability
of $(\nu_L,\, \nu_M,\, \nu_H)$ we need to know the mass hierarchy from
an external source, such as long-baseline neutrino oscillation
experiments (\cf, \Sec~\ref{sec:nova}).

\TABLE[!t]{
  \begin{tabular}{cccrr|rr|rr|rr}
      \hline
      \multicolumn{5}{c|}{Simulated scenario}
      & \multicolumn{6}{c}{Marginalized} \\
      No. & ($a$,$b$) & Hier. & $R$ & $S$
      & $\Delta \chi_0^2$ & $\sigma_0$
      & $\Delta \chi_1^2$ & $\sigma_1$
      & $\Delta \chi_2^2$ & $\sigma_2$ \\
      \hline 
      \#2 &  (0,1)  & NH & 0.33 & 2.72 & 2.6 & 1.6 &  3.6 & 1.9 &  3.6 & 1.9 \\
      \#2 &  (0,1)  & IH & 0.56 & 0.57 & 4.2 & 2.0 & 10.4 & 3.2 & 12.0 & 3.5 \\
      \#3 &  (1,--) & IH & 0.74 & 0.25 & 3.8 & 1.9 & 51.5 & 7.2 & 63.1 & 7.9 \\
      \#7 & (--,--) & NH & 0.21 & 4.67 & 0.0 & 0.2 & 13.4 & 3.7 & 13.4 & 3.7 \\
      \#7 & (--,--) & IH & 0.83 & 0.00 & 0.1 & 0.2 & 67.0 & 8.2 & 70.4 & 8.4 \\
      \hline
  \end{tabular}
  \caption{\label{tab:hier}%
    Identification of the neutrino mass hierarchy from astrophysical
    measurements only. The first four columns refer to the simulated
    scenario, hierarchy, and benchmark $R$ and $S$ values (marked in
    \figu{hierarchy}, lower row).  The last six columns represent the
    overall $\Delta \chi^2$ and $\sigma$ for the wrong hierarchy
    exclusion marginalized over all physics scenarios. In these
    columns, we distinguish $\Delta \chi_0^2$ and $\sigma_0$ for no
    special assumptions, $\Delta \chi_1^2$ and $\sigma_1$ for special
    case~1 in \Sec~\ref{sec:physics} (lightest state stable), and
    $\Delta \chi_2^2$ and $\sigma_2$ for special case~2 in
    \Sec~\ref{sec:physics} (no invisible states). We assume $L=100$
    muon tracks for this simulation of a pion beam source, and $S$ and
    $R$ to be measured. The $\chi^2$ from present solar, atmospheric,
    reactor, and accelerator data has been
    added~\cite{GonzalezGarcia:2007ib}.}}

For what concerns the \emph{mass hierarchy determination}, let us
first consider the generic case with all possible decay scenarios,
shown in the upper panels of \figu{hierarchy}. The fact that in a
given 123 scenario the allowed region for one hierarchy is always a
subregion of the one for the other hierarchy adds to the already
mentioned problem of the degeneracy between different scenarios, and
hence there is only a very limited portion of the parameter space
where the hierarchy can be determined unambiguously by an
astrophysical measurement. On the other hand, at the end of
\Sec~\ref{sec:physics} we discussed a number of special cases reducing
the number of possible decay scenarios and also restricting the
corresponding parameter space. These special cases can be either
motivated by specific decay models, or by phenomenological
observations. For example, one may assume that the lightest mass
eigenstate $\nu_L$ is stable (special case 1). This constraint implies
that only scenarios \#1, \#2, \#3, and \#7 are remaining, as shown in
the lower row of \figu{hierarchy}. If one further assumes that there
are no invisible states (special case 2), then the allowed branching
ratios become more restricted and the patterned regions disappear as
well. Clearly, the mass hierarchy can be now easily determined: for
example, $R \gtrsim 1.2$ would imply scenario \#7 and the inverted
hierarchy.  Note, however, that this interpretation of the
experimental result is no longer purely phenomenological, and is
intrinsically linked to the special assumption used.

We quantify this observation for several benchmark points in
\Tab~\ref{tab:hier}. The first four columns refer to the simulated
scenario, hierarchy, and benchmark $R$ and $S$ values (marked in
\figu{hierarchy}, lower panels). The last six columns represent the
overall $\Delta \chi^2$ and $\sigma$ for the rejection of the wrong
hierarchy, marginalized over all physics scenarios. Here we
distinguish $\Delta \chi_0^2$ and $\sigma_0$ for no special
assumptions, $\Delta \chi_1^2$ and $\sigma_1$ for special case~1
($\nu_L$ stable), and $\Delta \chi_2^2$ and $\sigma_2$ for special
case~2 (no invisible states). As can be seen, in the general case at
most a $2\sigma$ mass hierarchy determination is possible, even for
the relatively high luminosity considered here. On the other hand, in
special case~1 the mass hierarchy can be easily measured in most of
the discussed cases.

\section{Generalized source or diffuse flux}
\label{sec:gensource}

\FIGURE[!t]{
  \includegraphics[width=0.95\textwidth]{fig.cumulative.eps}
  \caption{\label{fig:efrac}%
    Allowed regions in the $(R,\, S)$ plane for the normal hierarchy
    and an unknown source. The left panel corresponds to the electron
    fraction $X_e$ marginalized over in the range $0 \le X_e \le 1/3$,
    while the right panel corresponds to it marginalized over the full
    range $0 \le X_e \le 1$. See main text for details.}}

Now what happens if we do not know anything about the source, such as
if we have a mixture of different sources, or even a diffuse flux? Can
we still learn something about physics? Let us assume a flavor
composition at source $(X_e,\, 1-X_e,\, 0)$, \ie, $X_e = f_e$ is the
electron (flavor) fraction, and there are no $\nu_\tau$'s produced.
Such a flavor composition might be observed for a combination of
different sources with different energy dependencies, or a diffuse
flux. In these cases, $X_e$ can be obtained as a (weighted) average of
the different $X_e^i$ from the different sources $i$. In the most
general case, we have $0 \le X_e \le 1$, where $X_e=0$ corresponds to
a $\mu$-damped source, $X_e \simeq 1/3$ to a pion beam source, and
$X_e=1$ to a source from neutron decays. Assuming that the neutrinos
are only produced by pion decays (and partly subsequent muon decays)
with an unknown energy dependence, we have $0 \lesssim X_e \lesssim
0.35$ from \Refs~\cite{Lipari:2007su, Pakvasa:2007dc} including
spectral effects. In general, any value of $X_e$ is possible, but only
one physics scenario will be realized if the decays are complete.
Furthermore, let us assume, for the sake of simplicity, that we know
the mass hierarchy from a different source.

\TABLE[!t]{
  \begin{tabular}{crr|rrrrrrr|rr} 
      \hline
      \multicolumn{3}{c}{Simulated scenario} & \multicolumn{8}{c}{Fit scenario $\Delta \chi^2$} \\
      No. & $R$ & $S$ &  \#1 & \#2 & \#3 & \#4 & \#5 & \#6 & \#7 & Any & $\sigma$ \\
      \hline 
      \multicolumn{11}{l}{\bf Simulated pion beam source, fit $\boldsymbol{0 \le X_e \le 1/3}$:
	($\bigtriangleup\square$)} \\
      \#1 & 0.49 & 1.07 &   -- &  1.0 &  0.0 &  26.4 & 220.3 &  0.6 &  55.3 &  0.0 & 0.1 \\
      \#2 & 0.38 & 2.19 & 21.0 &   -- &  0.4 &  73.1 & 432.1 & 20.0 &  20.6 &  0.4 & 0.6 \\
      \#3 & 0.38 & 1.47 &  5.1 &  0.2 &   -- &  47.4 & 274.0 &  7.7 &  32.7 &  0.2 & 0.4 \\
      \#4 & 0.69 & 0.41 &  1.9 & 25.6 &  0.1 &    -- &  43.7 & 15.4 & 115.7 &  0.1 & 0.3 \\
      \#5 & 0.83 & 0.00 & 40.2 & 74.4 & 27.5 &  24.0 &    -- & 62.9 & 153.5 & 24.0 & 4.9 \\
      \#6 & 0.59 & 1.04 &  0.9 &  1.7 &  0.7 &  12.8 & 106.2 &   -- &  48.2 &  0.7 & 0.8 \\
      \#7 & 0.21 & 4.67 & 77.6 & 11.1 & 22.3 & 138.5 & 516.2 & 54.8 &    -- & 11.1 & 3.3 \\
      \hline
      \multicolumn{11}{l}{\bf Simulated muon damped source, fit $\boldsymbol{0 \le X_e \le 1/3}$:
	(\raisebox{1.9pt}{$\bigtriangledown$}$\square$)} \\
      \#1 & 0.60 & 0.61 &   -- & 13.8 &  0.0 &   4.0 &  96.1 &  5.3 &  99.5 &  0.0 & 0.1 \\
      \#2 & 0.41 & 1.92 & 11.2 &  --  &  0.1 &  57.7 & 337.4 & 13.1 &  26.1 &  0.1 & 0.4 \\
      \#3 & 0.49 & 0.82 &  0.2 &  4.1 &   -- &  13.2 & 125.9 &  0.4 &  62.0 &  0.2 & 0.5 \\
      \#4 & 0.71 & 0.33 &  5.3 & 36.1 &  0.5 &    -- &  34.1 & 24.7 & 144.2 &  0.5 & 0.7 \\
      \#5 & 0.83 & 0.00 & 51.6 & 92.9 & 35.2 &  31.1 &    -- & 80.2 & 192.1 & 31.1 & 5.6 \\
      \#6 & 0.59 & 1.04 &  0.9 &  1.8 &  0.7 &  13.1 & 109.7 &   -- &  48.9 &  0.7 & 0.8 \\
      \#7 & 0.21 & 4.67 & 60.8 &  8.8 & 18.5 & 116.2 & 401.5 & 49.1 &    -- &  8.8 & 3.0 \\
      \hline
      \multicolumn{11}{l}{\bf Simulated neutron beam source, fit $\boldsymbol{0 \le X_e \le 1}$:
	(\raisebox{1.8pt}{$\scriptstyle\bigcirc$})} \\
      \#1=\#2 & 0.31 & 2.86 &  --  &   -- & 0.0 & 35.5 & 740.0 & 35.5 & 10.5 &  0.0 & 0.0 \\
      \#3=\#7 & 0.21 & 4.67 & 14.0 & 14.0 &  -- & 72.9 & 810.9 & 72.8 &   -- & 14.0 & 3.7 \\
      \#4=\#6 & 0.59 & 1.04 &  0.8 &  1.6 & 0.6 &   -- &  95.3 &   -- & 45.9 &  0.6 & 0.8 \\
      \hline
  \end{tabular}
  \caption{\label{tab:gensource}%
    Same as \Tab~\ref{tab:ex} for different sources, and $X_e$
    marginalized in the indicated ranges. That means that here the
    source is assumed to be unknown to some degree, or one measures a
    diffuse flux (superposition of sources). Here $L=100$ and a
    measurement of $R$ and $S$ is assumed. The different (simulated)
    benchmark points are marked in \figu{efrac}. The $\chi^2$ from
    present solar, atmospheric, reactor, and accelerator data has been
    added~\cite{GonzalezGarcia:2007ib}.}}

We show in \figu{efrac} the allowed regions for the observables $R$
and $S$ for the scenarios from \Tab~\ref{tab:phys} for the normal
hierarchy (99\% CL). Let us first of all assume that we do not know
anything about the source(s), \ie, $0 \le X_e \le 1$. Therefore, we
marginalize in the right panel of \figu{efrac} over $X_e$ in the full
range $0 \le X_e \le 1$, which means that the regions span the whole
range between muon damped and neutron beam source.\footnote{For
arbitrary marginalizations $0 \le X_e \le X_e^\text{max}$, and
arbritrary fixed $X_e$, see movies in \App~\ref{app:movies}.} As the
first observation, the scenario with only one final active stable
state remains unchanged in consistency with \equ{flrspecial}. As a
consequence, scenario~\#5 is still easy to identify. For the rest of
the scenarios there is relatively strong overlap, and only in rare
cases the scenarios might be identified. Nevertheless, many scenarios
can still be excluded. 

If we assume that only pion beam and muon damped sources (and mixtures
of these) contribute, we find the result in the left panel of
\figu{efrac}. Such an mixture might be measured for very limited
energy resolution, unknown source parameters of a specific source, or
a diffuse flux in a certain energy range. In this case, the result is
qualitatively not extremely different from the previous discussion.
For example, scenarios~\#5 and~\#7 are still relatively easy to
identify. In addition, the conclusions from the previous chapters
remain qualitatively unchanged. For a quantitative update, see
\Tab~\ref{tab:gensource}, which is similar to \Tab~\ref{tab:ex}, but
for different sources, and $X_e$ marginalized in the indicated ranges.
That means that here the source is assumed to be unknown to some
degree, or one measures a diffuse flux (superposition of sources).
Here $L=100$, and a simultaneous measurement of $R$ and $S$ is
assumed. The different (simulated) benchmark points are marked in
\figu{efrac}, where the first group corresponds to the triangles 
$\bigtriangleup$ and squares $\square$ the left panel (simulated
$X_e=1/3$), the second group to the triangles
\raisebox{1.9pt}{$\bigtriangledown$} and squares $\square$ in the left
panel (simulated $X_e=0$), and the third group to the circles
$\scriptstyle\bigcirc$ in the right panel (simulated
$X_e=1$).\footnote{Boxes are used if both points concide, which is the
case for all scenarios with only one stable mass eigenstate. In this
case, the observables to not depend on the flavor composition at the
source; \cf, \equ{flrspecial}.} It is now very interesting to compare
the first group in \Tab~\ref{tab:gensource} to the first group in
\Tab~\ref{tab:ex}, which are only different by the marginalization
over $X_e$. While in some cases the result does not change at all
(such as for simulated scenario \#1 and fit scenario \#5), the
sensitivity is in some cases completely destroyed (such as for
simulated scenario \#4 and fit scenario \#3). This can be easily
understood from \figu{efrac}, since the corresponding regions now
overlap each other. Similar results are obtained for the simulated
muon damped source in the middle row of \Tab~\ref{tab:gensource}. For
the neutron beam source in the last row of \Tab~\ref{tab:gensource}
there is, however, a qualitative difference: Since the mass eigenstate
$\nu_3$ is initially not populated for $X_e=1$ because we assume a
simulated $\stheta=0$ (and therefore $U_{e3}=0$; \cf, \equ{prob} and
\equ{flr}), the stability of $\nu_3$ is irrelevant, and models that
differ only in that stability are physically equivalent. This means
that the simulated models are paired, \ie, \#1=\#2, \#3=\#7, \#4=\#6,
\#5=\#8 (and this last case is irrelevant since nothing arrives at the
detector).\footnote{There can, however, be a difference in the fit
$\chi^2$ between two paired models, because we allow for $\stheta>0$
in the fit.} From \Tab~\ref{tab:gensource}, it is quite interesting
that in scenario \#3=\#7, any other qualitative case can be
significantly excluded even if the only assumption on the source is
that there are almost no $\nu_\tau$'s produced.

\section{Glashow resonance process as a third observable?}
\label{sec:glashow}

\FIGURE[!t]{
  \includegraphics[width=0.95\textwidth]{fig.chiral.eps}
  \caption{\label{fig:glashow}%
    Allowed regions at 99\% CL in the $(T,\, S)$ plane, for a pion
    beam source and a normal hierarchy. The left panel corresponds to
    $pp$ neutrino production, while the right panel corresponds to
    $p\gamma$ production. In the legend, the black and white disks
    correspond to stable and unstable mass eigenstates,
    respectively.}}

The Glashow resonance process $\bar{\nu}_e + e^- \to W^- \to
\text{anything}$ at around $6.3 \,
\text{PeV}$~\cite{Learned:1994wg} allows for
the detection of electron antineutrinos only. Therefore, we define $T
= \hphidet_{e^-} / \phidet_\mu$ as an additional observable. This is
the only observable which is sensitive to the production of pions (and
kaons) at the source by interactions of high energy protons with
photons (``$p\gamma$'') or protons (``$pp$'')~\cite{Anchordoqui:2004eb}.
In the $p\gamma$ process, mainly $\pi^+$ are produced through the $\Delta$
resonance, which means that the flavor composition at the source is
$(\hat{f}_e,\, \hat{f}_\mu,\, \hat{f}_\tau \,|\, \hat{f}_{\bar{e}},\,
\hat{f}_{\bar{\mu}},\, \hat{f}_{\bar{\tau}}) \simeq (\frac{1}{3},\,
\frac{1}{3},\, 0 \,|\, 0,\, \frac{1}{3},\, 0)$ (split up by neutrinos
and antineutrinos with $\sum \hat f_i = 1$). In the $pp$ process, a
nearly equal mix between $\pi^+$ and $\pi^-$ is produced, leading to
$(\hat{f}_e,\, \hat{f}_\mu,\, \hat{f}_\tau \,|\, \hat{f}_{\bar{e}},\,
\hat{f}_{\bar{\mu}},\, \hat{f}_{\bar{\tau}}) \simeq (\frac{1}{6},\,
\frac{1}{3},\, 0 \,|\, \frac{1}{6},\, \frac{1}{3},\, 0)$. If the
detector is CP-blind and there is no asymmetry between neutrinos and
antineutrinos, we can sum the flavor compositions of neutrinos and
antineutrinos at the source in order to obtain $(f_e,\, f_\mu,\,
f_\tau) = (\hat{f}_e + \hat{f}_{\bar{e}},\, \hat{f}_\mu +
\hat{f}_{\bar{\mu}},\, \hat{f}_\tau + \hat{f}_{\bar{\tau}}) \simeq
(\frac{1}{3},\, \frac{2}{3},\, 0)$ in both cases. Similarly, if the
muons are damped at higher energies, we have $(\hat{f}_e,\,
\hat{f}_\mu,\, \hat{f}_\tau \,|\, \hat{f}_{\bar{e}},\,
\hat{f}_{\bar{\mu}},\, \hat{f}_{\bar{\tau}}) \simeq (0,\, 1,\, 0 \,|\,
0,\, 0,\, 0)$ for $p\gamma$, and $(\hat{f}_e,\, \hat{f}_\mu,\,
\hat{f}_\tau \,|\, \hat{f}_{\bar{e}},\, \hat{f}_{\bar{\mu}},\,
\hat{f}_{\bar{\tau}}) \simeq (0,\, \frac{1}{2},\, 0 \,|\, 0,\,
\frac{1}{2},\, 0)$ for $pp$. In addition to the electron fraction
$X_e$ describing the fraction of electron neutrinos (and electron
antineutrinos), one can introduce a photon fraction $X_\gamma$
describing the fraction of neutrinos produced by $p\gamma$ processes.
In this case, the fraction $1-X_\gamma$ comes from $pp$ interactions.
If we assume that all electron neutrinos (antineutrinos) come from
antimuon (muon) decays, one can parameterize the source as
\begin{multline}
    \label{equ:allfrac}
    (\hat{f}_e,\, \hat{f}_\mu,\, \hat{f}_\tau \,|\,
    \hat{f}_{\bar{e}},\, \hat{f}_{\bar\mu}, \hat{f}_{\bar\tau}) =
    \\
    \bigg( \frac{1+X_\gamma}{2} X_e,\,
    \frac{1+X_\gamma-X_e-3 X_\gamma X_e}{2},\, 0 \,\bigg|\,
    \frac{1-X_\gamma}{2} X_e,\,
    \frac{1 - X_\gamma - X_e + 3 X_\gamma X_e}{2},\, 0 \bigg) \,.
\end{multline}
Conversely, the fraction of observed (useful) muon decays $0 \le \xi
\le 1$ is given by $\xi = X_e/(1-2 X_e)$. Note that this
parameterization can only describe the above decay chain, and is not
useful for $X_e>1/3$, such as for a neutron beam source.

We show in \figu{glashow} the observables $T$ and $S$ for a pion beam
source ($X_e=1/3$) and $pp$ (left panel, $X_\gamma=0$) versus
$p\gamma$ interactions (right panel, $X_\gamma=1$).\footnote{For
arbitrary $X_\gamma$, see movies in \App~\ref{app:movies}.} Note that
neither $R$ nor $S$ depend on $X_\gamma$ since the neutrino and
antineutrino rates are added. Since the regions do not collapse to
thin curves, \ie, $T$ is a well-defined function $T(S)$, there is
obviously additional useful information in $T$. However, in the $pp$
case, the different scenarios cluster along the diagonal and overlap
each other in the same fashion as for $S$. Therefore, for scenario
identification, $T$ may not provide much new information. For the
$p\gamma$ case, however, there is obviously new information. For
example, if $(T,\, S) = (0.1,\, 1)$ is measured, scenario \#3 can be
uniquely established, as well as the $p\gamma$ source can be
identified. If only $R$ and $S$ were used, it would be fully contained
in clusters \#2 and \#6 (\cf, \figu{flratios}, upper right panel).
Note that one can also establish the $pp$ source in some cases. For
example, if $T \gtrsim 4$ is observed, scenario \#7 together with a
$pp$ source has to be realized.

\section{Synergies with terrestrial neutrino oscillation experiments}
\label{sec:synergies}

In this section we discuss the synergies with the terrestrial neutrino
oscillation experiments. We focus on two decay scenarios, which are,
in wide ranges of the parameters, relatively easy to identify:
scenario \$5 (only $\nu_3$ stable) and scenario \$7 (only $\nu_1$
stable). Since in both cases there is only one stable mass eigenstate,
the observables depend on the mixing matrix elements only (\cf,
\equ{flrspecial}). In particular, there is no dependence on the flavor
composition at the source, branching ratios, and mass hierarchy. This
means that there is also no mass hierarchy information from the 
astrophysical neutrino source(s). As discussed in \figu{params}, $R$
varies strongly with $\deltacp$ in scenario \$7 (see also
\Ref~\cite{Beacom:2003zg}), whereas $R$ varies strongly with
$\theta_{23}$ in scenario \$5. Therefore, we use these two scenarios
in combination with terrestrial measurements, and compare them to the
standard oscillation result scenario (no decays) or the \emph{same}
luminosity of the source. As for the time scale, we choose the next
generations of reactor and long-baseline experiments. Namely we use
Double Chooz, MINOS, and NO$\nu$A as examples. In addition, we assume
that there are no decay effects observed in terrestrial experiments,
\ie, neutrino decay is visible only over astronomical distances. For
details on the statistical simulation, see \App~\ref{app:statistics}.

\subsection{Can Double Chooz plus neutrino telescope measure
  $\boldsymbol{\deltacp}$?}

\FIGURE[!t]{
  \includegraphics[width=\textwidth]{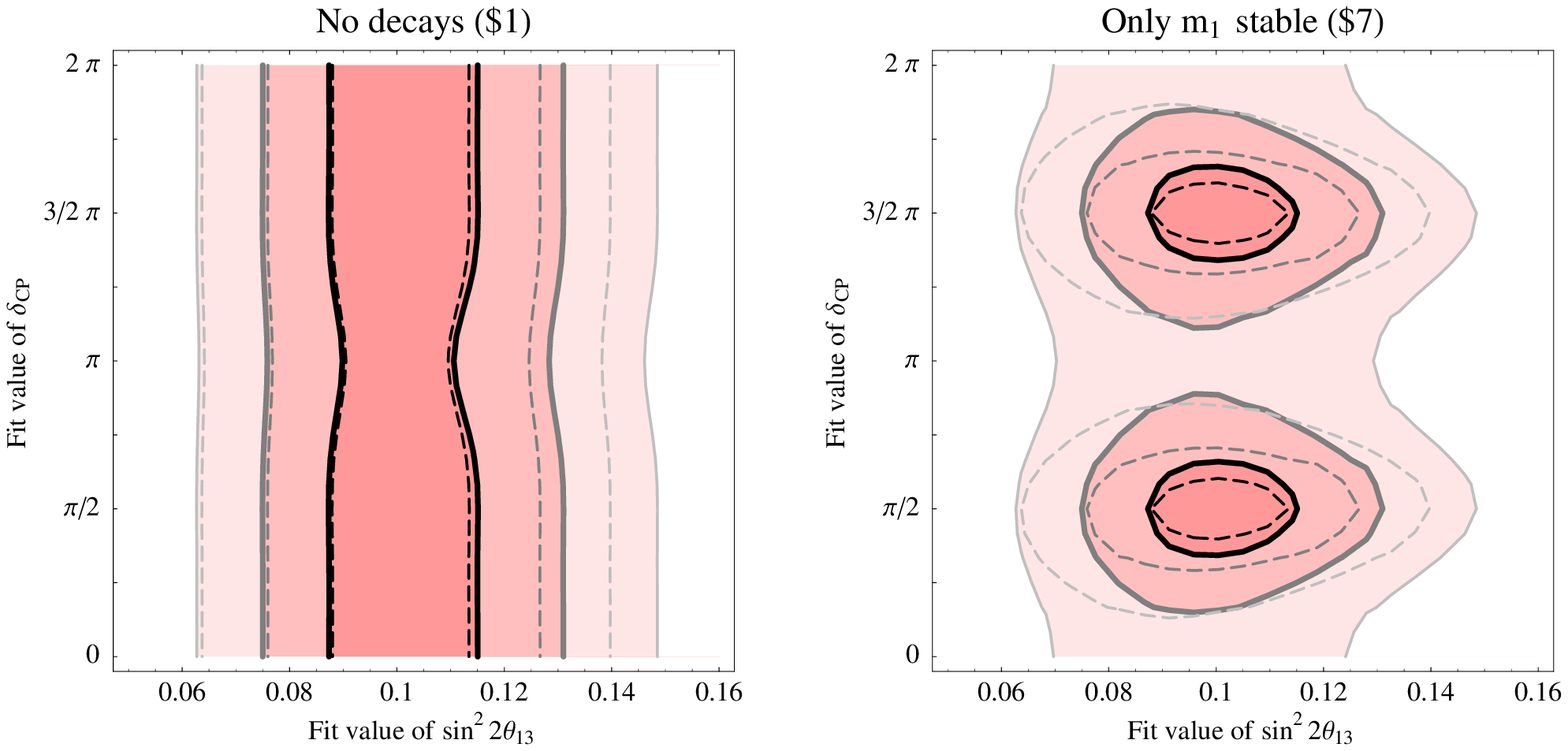}
  \caption{\label{fig:sourcecomp}%
    Comparison between physics scenarios \$1 (no decay, left column),
    and \$7 ($\nu_1$ stable, right column), for Double Chooz plus an
    astrophysical source.  Here a pion beam source $L = 100$ tracks as
    normalized luminosity for both sources is assumed, \ie, the left
    and right panels correspond to the same source luminosity. In
    addition, it is assumed that only $R$ can be measured. The
    contours correspond to $1\sigma$, $2\sigma$, $3\sigma$ (1 d.o.f.).
    The dashed curves are for fixing the other oscillation parameters.
    The current best-fit values and parameter errors are taken from
    \Ref~\cite{GonzalezGarcia:2007ib}. The used simulated values are
    $\stheta=0.1$, $\deltacp=\pi/2$, and a normal hierarchy.}}

As in was pointed out in \Refs~\cite{Winter:2006ce, Blum:2007ie},
flavor ratio measurements might allow a measurement of $\deltacp$
already in combination with Double Chooz. However, for a pion beam
source, which may be the most common one, the dependence of $R$ and
$S$ on $\deltacp$ and the other oscillation parameters in the standard
no-decay scenario is very moderate (\cf, \figu{flratios}). In this
case, knowledge from different sources, high statistics, and the use
of different observables is necessary to obtain useful information on
$\deltacp$. However, if neutrinos decay there can be a relatively
strong dependence on $\deltacp$, depending on the specific
scenario~\cite{Beacom:2003zg}. We demonstrate this effect
quantitatively for a three year Double Chooz measurement and a
relatively large $\stheta$ in \figu{sourcecomp}. In this figure, the
precision in $\stheta$-$\deltacp$ is shown for maximal CP violation
implemented by nature. For the astrophysical source, we only assume a
pion beam source producing 100 muon tracks for the standard scenario,
and we only measure the observable $R$. The left and right panels
correspond to the same source luminosity. Obviously, if neutrinos are
stable, there will be hardly any information on $\deltacp$. However,
if only $\nu_1$ is stable, even a $2\sigma$ CP violation measurement
might be possible (if the uncertainties on the other oscillation
parameters can be further reduced, even $3\sigma$ -- see dashed
curves).

\subsection{Octant determination with terrestrial experiments plus neutrino
  telescope}

\FIGURE[!t]{
  \includegraphics[width=\textwidth]{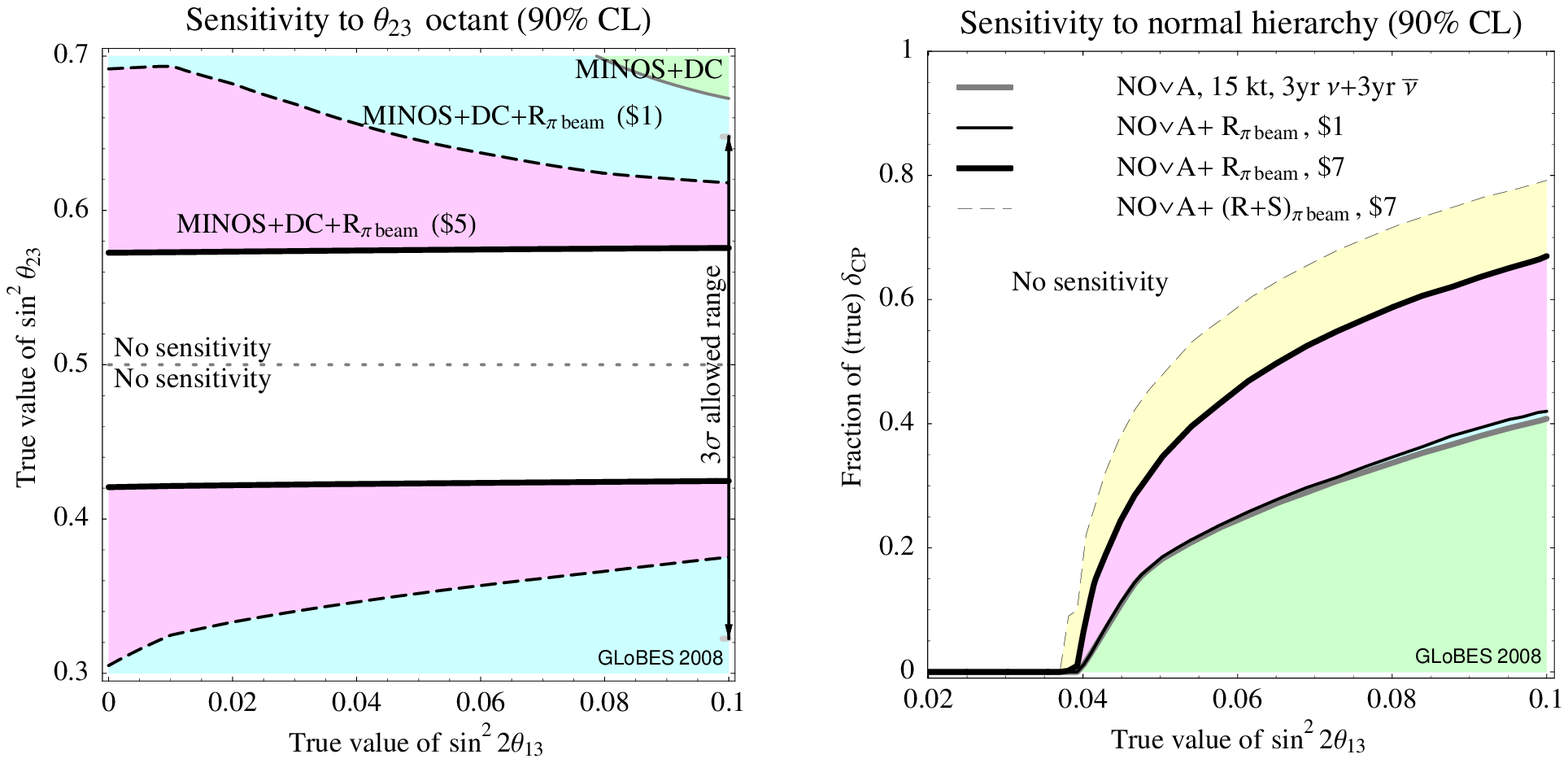}
  \caption{\label{fig:novamh}%
    Left panel: Sensitivity to the $\theta_{23}$ octant as a function of
    the true $\stheta$ and true $\sin^2\theta_{23}$, where sensitivity is given at
    the 90\% CL above/below the curves (as illustrated by the shadings). 
    The different curves represent MINOS plus Double Chooz (DC) alone (thin gray curve), MINOS plus Double Chooz
    plus astrophysical information in scenario \$1 (dashed curves), and 
    MINOS plus Double Chooz plus astrophysical information in scenario \$5 (thick solid curves).
    Right panel: Sensitivity to the normal mass
    hierarchy as a function of the true $\stheta$ and true $\deltacp$ (stacked to
    the ``Fraction of $\deltacp$''), where sensitvity is given at the 90\% CL below the
    curves. The different curves
    represent NO$\nu$A alone (thick gray curve), NO$\nu$A plus 
    astrophysical in scenario \$1 (thin black curve),  NO$\nu$A plus 
    astrophysical in scenario \$7 (thick black curve), and NO$\nu$A plus 
    astrophysical in scenario \$7 with both $R$ and $S$ as observables (dashed curve).
    In both plots,  $L = 100$ muon tracks were assumed for the flux normalization, 
    only $R$ was used as an observable (unless stated otherwise), and a normal
    hierarchy was simulated. The octant plot does not include the
    mixed (octant and sign) degeneracy. For details on the simulation,
    see \App~\ref{app:statistics}.}}

We define sensitivity to the $\theta_{23}$ octant if the wrong octant
can be excluded at the 90\% CL. Quite naturally, there will be no
sensitivity if $\sin^2 \theta_{23}$ is close to $0.5$, and there will
be no sensitivity if the observables depend only on $\sin^2 2 \theta_{23}$,
as it applies to the disappearane channel in long-baseline experiments.
The improvement of the octant measurement using astrophysical
neutrinos was discussed in \Ref~\cite{Winter:2006ce} for no decays and
a combination of terrestrial experiments. Here we focus on a shorter
time scale. We assume that we have information from Double Chooz and
MINOS. We show in \figu{novamh}, left panel, the sensitivity to the 
$\theta_{23}$ octant as a function of the true $\stheta$ and true 
$\sin^2\theta_{23}$, where sensitivity is given at
    the 90\% CL above/below the curves (as illustrated by the shadings). 
    The different curves represent MINOS plus Double Chooz alone (thin gray curve), MINOS plus Double Chooz
    plus astrophysical information in scenario \$1 (dashed curves), and 
    MINOS plus Double Chooz plus astrophysical information in scenario \$5 (thick solid curves).
We can read off this figure that there
is no sensitivity to the octant in the $3\sigma$ currently allowed
$\theta_{23}$ range (marked in the figure). If $R$ is measured for
stable neutrinos (\$1) and for a pion beam source, the potential
substantially improves compared to no astrophysical information, and
includes wide region of the currently allowed range. Typically, it depends
on the true $\stheta$. However, if only
$\nu_3$ is stable (\$5), we can read off from \equ{flrspecial} that
\begin{equation}
    R = \frac{\sin^2 \theta_{23} \cos^2 \theta_{13}}
    {1 - \sin^2 \theta_{23} \cos^2 \theta_{13}}
    \simeq \tan^2 \theta_{23} \,.
\end{equation}
This implies that $\theta_{23}$ can be measured almost without
parameter correlation by the neutrino telescope. In the left panel of
\figu{novamh} we show the excellent precision compared to scenario \$1
for the same source luminosity and observable, where the sensitivity
mainly comes from the astrophysical source.

\subsection{Mass hierarchy determination with NO$\nu$A plus
  astrophysical}
\label{sec:nova}

For terrestrial long-baseline experiments, the mass hierarchy
degeneracy~\cite{Minakata:2001qm}, which determines the mass hierarchy
measurement, is, in general, located at a different value of
$\deltacp$ than the original solution. In addition, it moves in the
$\deltacp$ direction as a function of the true $\stheta$ (\cf, \Fig~4
in \Ref~\cite{Winter:2006ce} for NO$\nu$A). Since astrophysical
neutrino sources are sensitive to $\cos \deltacp$, whereas first
generation superbeams operated close to the oscillation maximum are
mainly sensitive to $\sin \deltacp$, the knowledge from the
astrophysical source can improve the mass hierarchy measurement at the
terrestrial experiments~\cite{Winter:2006ce}. This effect is largest
in scenarios where the dependence of the observables on $\deltacp$ is
strongest. We illustrate this behavior in the right panel of
\figu{novamh}, where we show the sensitivity to the normal mass
    hierarchy as a function of the true $\stheta$ and true $\deltacp$ (stacked to
    the ``Fraction of $\deltacp$''). The different curves
    represent NO$\nu$A alone (thick gray curve), NO$\nu$A plus 
    astrophysical $R$ in scenario \$1 (thin black curve),  NO$\nu$A plus 
    astrophysical $R$ in scenario \$7 (thick black curve), and NO$\nu$A plus 
    astrophysical $R$ and $S$ in scenario \$7 (dashed curve).
We compare our results for scenarios \$1 (no
decays) and \$7 (only $\nu_1$ stable) for a pion beam source. Note
that, as explained before, there is no sensitivity at all to the mass
hierarchy from the astrophysical source alone. As can be seen from
this figure, there is almost no improvement for stable neutrinos
(\$1), where the mass hierarchy can be determined for up to 
(depending in $\stheta$) 40\% of all true $\deltacp$. However, if 
only $\nu_1$ is stable (\$7), 
a measurement of $R$
helps significantly, and the mass hierarchy can already be determined
for up to about 65\% of all true $\deltacp$. The additional information on $S$,
as shown as the dashed curve, helps even more.

\section{Summary and conclusions}
\label{sec:summary}

In this study, we have discussed the identification of different decay
and oscillation scenarios at neutrino telescopes. Furthermore, we have
studied the measurement of the physics parameters within these
scenarios by a neutrino telescope alone, and in combination with
future terrestrial experiments. We have taken into account the present
knowledge of the oscillation parameters from a global fit of current
solar, atmospheric, reactor, and accelerator data, we have
statistically quantified the information from the astrophysical
sources, and we have performed a complete simulation of future
terrestrial experiments. For the observables, we have mainly focused
on the muon track to shower ratio and the electromagnetic to hadronic
shower ratio, but we have also discussed the Glashow resonance
process. We have performed a complete classification of effective
decay scenarios and the corresponding branching ratios for complete
decays, accounting also for possible invisible states. We have
demonstrated that, depending on the physics scenario implemented by
nature, the identification of the scenario can be unique or ambiguous.
For example, if only $\nu_1$ or $\nu_3$ is stable (either of which can
be the lightest depending on the hierarchy), the physics scenario can
be easily identified. In the standard oscillation case, however, only
specific scenarios can be excluded.

As far as the impact of the mass hierarchy is concerned, we have
demonstrated that one may be able to establish which of the $(\nu_1,\,
\nu_2,\, \nu_3)$ mass eigenstates are stable, but not their mass
ordering, which determines the branching ratios (since only heavier
states can decay into lighter ones). For example, without external
mass hierarchy measurement such as from superbeams, one can in
principle determine whether $\nu_1$ is stable or not, but not if
$\nu_1$ is the lightest or middle mass eigenstate. This implies that a
generic mass hierarchy identification is only possible in very small
corners of the parameter space. However, if one imposes some
model-dependent constraints then the mass hierarchy can be easily
determined in most cases from astrophysical neutrinos alone. One
possible such constraint is the assumption that the lightest neutrino
mass eigenstate is stable. 

We have also studied the impact of flavor composition uncertainties at
the source or the use of diffuse fluxes. For example, we have
demonstrated that if no more than one active neutrino mass eigenstate
is stable, there is no dependence of the observables on the flavor
composition at the source. In order to study diffuse fluxes, we have
marginalized the electron fraction at the source, which means that
have taken into account arbitrary combinations of muon damped and pion
beam sources. In this case, the physics scenario identification
becomes quantitatively more difficult, but the qualitative conclusions
still hold. Even if one allows for arbitrary production of $\nu_e$ and
$\nu_\mu$ neutrinos at the source, some physics scenarios can still be
established.

For what concerns the determination of the neutrino parameters in
particular decay scenarios, we have chosen a number of examples in
order to demonstrate the impact of an astrophysical measurement for
future long-baseline and reactor experiments. For example, if a
neutrino telescope measures the track to shower ratio from a pion beam
source, then Double Chooz might be the first experiment to establish
CP violation if $\nu_1$ is the only stable state, whereas we have not
found any CP violation sensitivity if all the neutrinos are stable. As
another example, we have demonstrated that there is some sensitivity
to the $\theta_{23}$ octant if MINOS and Double Chooz are combined
with astrophysical data even if all neutrino mass eigenstates are
stable, but if only $\nu_3$ is stable there will be direct octant
sensitivity from the astrophysical source alone. We have also
illustrated how the mass hierarchy sensitivity at NO$\nu$A would be
enhanced by an astrophysical neutrino measurement if only $\nu_1$ is
stable. For large $\stheta$, the fraction of $\deltacp$, for which the
hierarchy can be determined, could increase from 40\% up to 80\%. Note
that this mass hierarchy determination is independent of any
model-dependent assumptions on the decay scenarios. 
 
We conclude that an observation of astrophysical neutrinos at a
neutrino telescope would be an important test of the oscillation and
decay neutrino properties.. While it is difficult to obtain
information on the neutrino lifetime without a distance measurement of
the source, complete decay scenarios can in many cases be easily
identified even if one takes into account the current measurement
precisions of the oscillation parameters and uncertainties of the
flavor composition at the source. Especially if neutrinos decay, the
combination with terrestrial neutrino experiments may lead to early
and surprising results even for the standard oscillation parameter
measurements. An important prerequisite for such conclusions will be
the flavor identification in the detector. 

\acknowledgments

We are grateful to Raj Gandhi for useful discussions.  WW would like
to thank the theory group at UAM for their hospitality during his
visit.
MM is supported by MCI through the Ram\'on y Cajal program and through
the national project FPA2006-01105, by the Comunidad Aut\'onoma de
Madrid through the HEPHACOS project P-ESP-00346, and by the European
Union through the ENTApP network of the ILIAS project
RII3-CT-2004-506222.
WW acknowledges support from the Emmy Noether program of Deutsche
Forschungsgemeinschaft.

\appendix

\section{Statistical method and simulation}
\label{app:statistics}

A detailed description of our simulation of present solar,
atmospheric, reactor and accelerator neutrino experiment can be found
in \Ref~\cite{GonzalezGarcia:2007ib}, from which we also take the
current best-fit values and allowed parameter ranges. For the MINOS
simulation, we follow \Ref~\cite{Huber:2004ug} with a total luminosity
of $5 \, \text{yr} \times 3.7 \cdot 10^{20} \, \text{pot} / \text{yr}$
and a $5.4 \, \text{kt}$ magnetized iron
calorimeter~\cite{Ables:1995wq} (the unit ``pot/yr'' refers to
``protons on target per year''). For Double
Chooz~\cite{Ardellier:2004ui, Ardellier:2006mn}, we use the simulation
from \Refs~\cite{Huber:2006vr, Huber:2007ji} with 1.5 years of data
taking with far detector only, followed by 1.5 years with both
detectors. For NO$\nu$A, we use the simulation from
\Refs~\cite{Huber:2002rs, Winter:2006ce} updated to the numbers from
\Ref~\cite{Ayres:2004js} and a $15 \, \text{kt}$ detector mass. The
future reactor and long-baseline experiments are simulated with the
GLoBES software~\cite{Huber:2004ka, Huber:2007ji}.

We define an astrophysical $\chi^2_\text{astro}$ to be added to the
GLoBES software or to present experiments as
\begin{equation}
    \label{equ:chi2}
    \chi^2_\text{astro} = \min_\xi \left\lbrace
    2 \sum_{i=1}^n \left[ T_i(\xi) - O_i
    + O_i \ln\frac{O_i}{T_i(\xi)} \right]
    + \left( \frac{\xi-1}{\sigma_\xi} \right)^2 \right\rbrace \,.
\end{equation}
Here $T_i$ corresponds to the theoretical (fit) rate and $O_i$ to the
observed (true) rate. The index $i$ runs over all event types, such as
muon tracks, showers, double-bang, \etc; $\xi$ is a source
type-dependent unknown (free) flux normalization parameter to be
marginalized over, and $\sigma_\xi$ its error. Since we are not
assuming any prior knowledge of the neutrino flux expected from the
observed source, we conservatively set $\sigma_\xi \to \infty$.

We consider two different cases for $T$ and $O$. If only $R$ is
measured we set $n=2$ and
\begin{equation} \begin{aligned}
    T_1 &= \xi N_\mu^\text{fit} \,, & \qquad
    T_2 &= \xi \left( N_e^\text{fit} + N_\tau^\text{fit} \right) \,,
    \\
    O_1 &= N_\mu^\text{true} \,, & \qquad
    O_2 &= N_e^\text{true} + N_\tau^\text{true} \,,
\end{aligned} \end{equation}
where $N_\beta$ is the (total) number of events for flavor
$\nu_\beta$. If both $R$ and $S$ are measured we set $n=3$ and
\begin{equation} \begin{aligned}
    T_1 &= \xi N_\mu^\text{fit} \,, & \qquad
    T_2 &= \xi N_e^\text{fit} \,, & \qquad
    T_3 &= \xi N_\tau^\text{fit} \,,
    \\ 
    O_1 &= N_\mu^\text{true} \,, & \qquad
    O_2 &= N_e^\text{true} \,, & \qquad
    O_3 &= N_\tau^\text{true} \,.
\end{aligned} \end{equation}
The event rate for the flavor $\nu_\beta$ in the detector is given as
\begin{equation}
    N_\beta = \phi \, \hat{\epsilon}_\beta
    \sum_{\alpha = 1}^3 \, f_\alpha \, P_{\alpha \beta}^{(k)}
\end{equation}
where $f_\alpha$ denotes the fraction of neutrinos produced as flavor
$\nu_\alpha$ at the source, $\hat{\epsilon}_\beta \equiv
\epsilon_\beta / \epsilon_\mu$ is a relative efficiency compared to
the muon track detection efficiency, $\phi$ corresponds to a
normalized luminosity at the detector, and $(k)$ refers to the decay
scenario in \Tab~\ref{tab:phys} (the probability is described by
\equ{prob}). In order to compare different physics scenarios for the
same source flux, we normalize to a number of muon tracks $L$ observed
in the detector for the standard oscillation scenario \#1 and
$\theta_{13}=0$, \ie,
\begin{equation}
    \phi = \frac{L}{\sum\limits_{\alpha = 1}^3
      f_\alpha P_{\alpha\mu}^{(1)}} \,.
\end{equation}
This normalization does not depend on the physics scenario. Therefore,
it allows to compare different physics scenarios for the same source
flux, and one can identify the physics in which one can most
efficiently measure the target parameter. In addition, the number of
observed muon tracks in the standard scenario is a quite intuitive
one. For a flux close to the Waxmann-Bahcall bound, one may expect $L
= \mathcal{O}(100)$ muon tracks in about eight years~\cite{Ahrens:2003ix}.

For the sake of simplicity, we assume that $\hat{\epsilon}_e =
\hat{\epsilon}_\tau = 1$. In a more realistic simulation one would
probably have $\hat{\epsilon}_e,\, \hat{\epsilon}_\tau \ll 1$, since
the detector is sensitive to partially contained muon track events
generated out of the fiducial volume, and the energy threshold for
$\mu$ events is lower~\cite{Beacom:2003nh}. Our assumption corresponds
to choosing appropriate cuts such that $\hat{\epsilon}_e \simeq
\hat{\epsilon}_\tau \simeq 1$, \ie, muon tracks and the other event
types are detected with similar efficiencies. For the case of $S$,
that of course implies relatively low event rates. In addition, we
assume a background-free environment. Backgrounds could be easily
included in our treatment, but they strongly depend on the source
type, energy range, \etc, whereas we want our simulation to be as much
source-independent as possible. Finally, we neglect flavor
identification uncertainties, whose impact on the determination of the
parameter $R$ has been estimated in \Ref~\cite{Beacom:2003nh} to be of
the order of 20\% -- whereas for the parameters $S$ and $T$ it is
considerably worse.

Although our approximations are likely to have a non-negligible impact
on our \emph{quantitative} results, it is quite easy to understand
their implications at \emph{qualitative} level. For example, the
uncertainties in $R$ and $S$ due to flavor identification can be
accounted for in \Figs~\ref{fig:flratios}--\ref{fig:glashow} by
describing the experimental results as extended ellipses rather than
points: degeneracies among models will then arise when more than one
region overlap with the experimental ellipse. Also, the main effects
of backgrounds and systematic uncertainties is to reduce the
significance of the fit, hence they are quite similar to a reduction
in statistics, which we have explicitly illustrated in
\Tab~\ref{tab:ex}. Therefore, we believe that despite our
simplications the main conclusions of our work are solid.

\section{Movies}
\label{app:movies}

A number of movies corresponding to \Sec~\ref{sec:physics},
\Sec~\ref{sec:gensource}, and \Sec~\ref{sec:glashow} can be found at
\Ref~\cite{AstroMovies}.

The first type of movies uses $X_e$ as a free parameter, \ie, the time
parameter, and corresponds to \figu{flratios}. The flavor composition
at the source is assumed to be $(f_e,\, f_\mu,\, f_\tau) = (X_e,\,
1-X_e,\, 0)$. Note that $X_e=0$ corresponds to a muon damped source,
$X_e=1/3$ to a pion beam source, and $X_e=1$ to a neutron beam source.
The movie shows the normal hierarchy (left) and the inverted hierarchy
(right). Movie versions corresponding to all rows in \figu{flratios}
are available.

The second type of movies is similar to the first type, but $X_e$ is
marginalized over in the range $0 \le X_e \le X_e^\text{max}$, and
$X_e^\text{max}$ is the time frame parameter. This movie corresponds
to \figu{efrac}, but movie versions corresponding to the different
rows in \figu{flratios} are available.

The third type of movies shows all observable pairs as a function of
the photon fraction $X_\gamma$ for a pion or muon damped source
(versions for the different mass hierarchies are available as well).
The photon fraction quantifies fraction of neutrinos is produced by
$p\gamma$ versus $pp$ processes. For more details, see
\Sec~\ref{sec:glashow}.


\providecommand{\href}[2]{#2}\begingroup\raggedright\endgroup

\end{document}